# Balancing Microservices and Monolithic Architectures

**Venkata Krishna Chaitanya Nuthalapati**
Software Development Engineer | Independent Researcher, Distributed Systems and DevSecOps

*Enterprise software teams face a fundamental architectural choice: build a single, unified application or decompose functionality into independently deployable services. This article examines both patterns—analyzing their technical benefits, documented drawbacks, and high-profile industry case studies—to help practitioners choose, combine, and evolve these two approaches in a principled way.*

## INTRODUCTION

Monolithic systems co-locate all components in one codebase, simplifying early development and deployment. Microservices architectures decompose an application into many loosely coupled services, each owning its data and independently deployable. As Felisberto notes [1], every architectural decision involves trade-offs, and the cloud era has both accelerated the adoption of distributed systems and triggered a notable reversal toward monoliths driven by concerns over cost, complexity, and performance.

According to the Cloud Native Computing Foundation (CNCF) 2024 Annual Survey, 80 percent of respondents now run Kubernetes in production—up from 66 percent in 2023—reflecting the growing prevalence of container-based, microservices-oriented deployments [3]. At the same time, Su, Li, and Taibi's multivocal literature review [6] documents a counter-trend: organizations are consolidating microservices back into monoliths, citing cost, complexity, scalability, performance, and organizational misalignment as reasons for reversal.

This article reviews both patterns for practitioners who must make or evaluate these decisions. We draw on recent peer-reviewed research and publicly documented industry case studies, and we offer concrete guidance for balancing the two styles in practice. Table I summarizes the principal trade-offs developed throughout.

## BACKGROUND: MONOLITH VS. MICROSERVICES

A monolithic architecture is a single, self-contained application in which all components share one codebase and one deployment unit [2]. This model simplifies early development: a team applies changes to one system, often backed by a shared database. A microservices architecture, by contrast, decomposes functionality into many services. Each service owns its data, can be built and deployed independently, and communicates with peers over well-defined Application Programming Interfaces (APIs) [2]. Figure 1 contrasts the two structures.

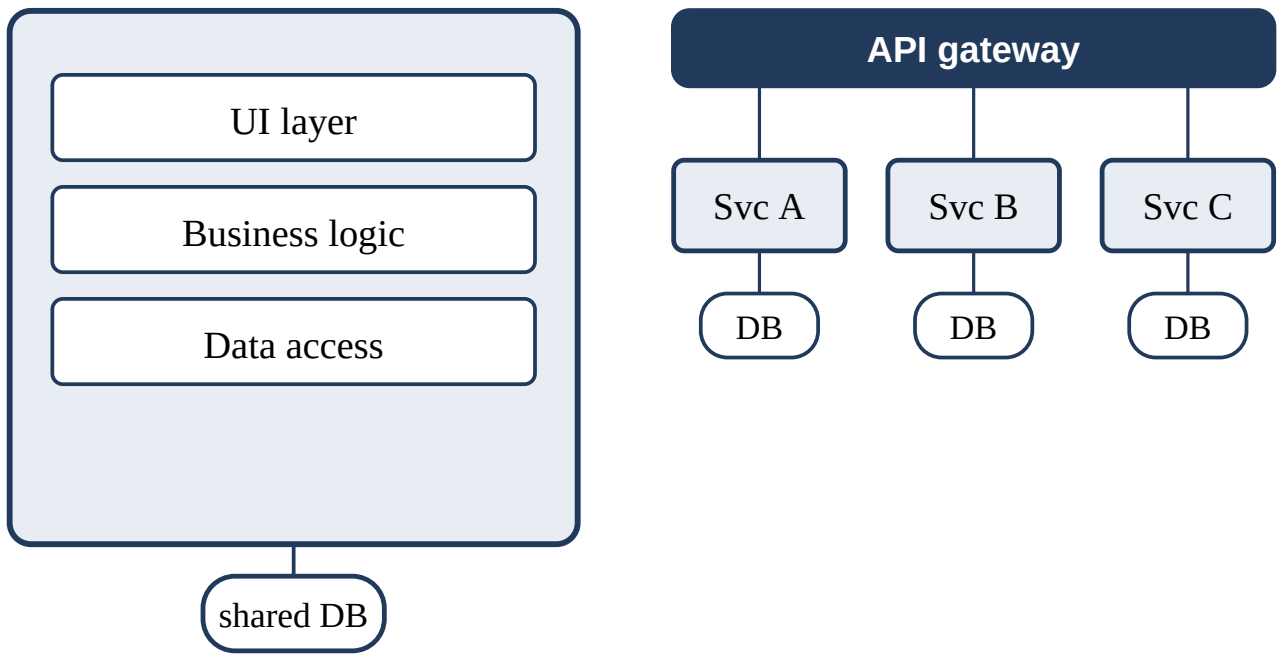

*Figure 1. Monolithic architecture (one deployable unit over a shared database) versus microservices (independently deployable services, each owning its data, communicating through an API gateway).*

Netflix illustrates the appeal of microservices at scale. After a catastrophic database corruption in August 2008 caused roughly three days of DVD-shipping downtime, Netflix began a seven-year migration of its monolith to Amazon Web Services (AWS). The company reported the migration complete in January 2016, having decomposed its monolith into hundreds of microservices running entirely in the cloud [4]. Atlassian's engineering team notes that Netflix now operates more than a thousand microservices supporting separate parts of the platform [2]. The transformation required years of coordinated engineering effort and a portfolio of purpose-built tooling that Netflix subsequently open-sourced—Conductor for workflow orchestration [21], Hystrix for latency and fault tolerance [22], and Eureka for service discovery [23].

Di Francesco, Lago, and Malavolta's industrial survey of microservices migrations highlights that neither style is universally superior; the right choice depends on an organization's scale, business domain, and operational maturity [5]. Felisberto's trade-off analysis [1] and Su et al.'s literature review [6] both confirm that the pendulum swings: organizations that over-decompose their systems have reversed course, merging services back to reduce cost and operational overhead.

## TECHNICAL ADVANTAGES OF MICROSERVICES

When requirements justify the added complexity, microservices deliver concrete engineering benefits.

### *Independent deployment and scalability.*

Each service scales or updates without touching others. Atlassian's "Vertigo" project demonstrates this at scale: while migrating roughly 100,000 Jira and Confluence instances to a multi-tenant AWS platform, Atlassian decomposed its monoliths into microservices—an effort its head of platform describes as the largest engineering undertaking in the company's history [24]. Atlassian's engineering team reports growing from about 15 microservices in January 2016 to more than 1,300, letting teams ship features to a single service without coordinating a monolithic release [2]. Blinowski, Ojdowska, and Przybylek's empirical study confirms that microservices achieve meaningfully higher throughput under high, concurrent load [7].

### *Technology flexibility.*

Teams select the language or framework best suited to each service's requirements. Atlassian reports that this polyglot capability gives teams the freedom to choose the tools they prefer [2], accelerating development when each team can leverage a familiar stack.

### *Fault isolation and reliability.*

A failure in one service need not propagate to the entire system. Atlassian notes that targeted deployments let teams update a specific service without risking the entire application [2]. Newman documents circuit-breaker and bulkhead patterns that microservices make tractable at service boundaries [8].

### *Team autonomy.*

Mapping services to small, cross-functional "two-pizza teams"—a concept originating with Amazon's early-2000s internal reorganization [9]—reduces coordination overhead. Atlassian engineers reported being more autonomous and able to build and deploy their own changes [2]. Skelton and Pais note that team boundaries and service boundaries co-evolve; neither fully determines the other [10].

## DRAWBACKS AND TRADE-OFFS

Microservices introduce significant technical and operational challenges that practitioners must address honestly.

### *Architectural complexity.*

Decomposing an application into many services means managing many codebases, repositories, databases, and APIs. Atlassian warns of "development sprawl"—the unmanaged proliferation of services across teams [2]. Di Francesco et al.'s migration survey found that the distributed nature and heterogeneous technology stacks of microservices introduce complexity that frequently surpasses that of a comparable monolith [5].

### *Infrastructure costs.*

Each microservice requires its own runtime environment, Continuous Integration/Continuous Delivery (CI/CD) pipelines, monitoring instrumentation, logging infrastructure, and security controls. Blinowski, Ojdowska, and Przybylek's IEEE Access evaluation found that on a single machine a monolith outperforms an equivalent microservice deployment, and that for systems without thousands of concurrent users—those that can still scale vertically—the efficiency gains expected from decomposition may prove illusory [7]. Su, Li, and Taibi's review identifies cost as among the five principal reasons organizations have reversed microservices migrations, alongside complexity, scalability, performance, and organizational misalignment [6].

### *Testing complexity.*

Microservices shift validation from unit-level to complex integration scenarios. Gregor, Hentschel, Kastner, and Pretschner's taxonomy of integration faults—published at the 2025 IEEE International Conference on Software Testing, Verification and Validation (ICST)—shows that communication failures, data-format mismatches, and API version incompatibilities form a distinct class of defects that unit tests cannot catch [12]. Signadot's analysis confirms the practical difficulty: shared staging environments degrade when multiple service versions run simultaneously, often forcing teams toward manual verification [13]. Lehva, Makitalo, and Mikkonen demonstrate that consumer-driven contract testing—as provided by the Pact framework—addresses these gaps by letting each service verify its interfaces against recorded consumer expectations [14].

### *Security and observability overhead.*

In a monolith, components share a common session and trust boundary. In a microservices architecture, every inter-service call requires authentication—typically via JSON Web Tokens (JWTs)—mutual Transport Layer Security (mTLS), and audit logging, multiplying the compliance surface [1]. Li et al.'s survey of service mesh technologies—such as Istio and Linkerd—describes how organizations address this overhead by routing all inter-service traffic through a sidecar proxy that enforces policy uniformly [15]. Sharma and Singh's IEEE Software study confirms that adopting a service mesh significantly reduces per-service security implementation burden, though it introduces new complexity in mesh configuration and certificate management [16].

Beyer, Jones, Petoff, and Murphy's foundational Site Reliability Engineering text notes that operational toil scales with service count [17]; in practice, tracing a single user request across dozens of independent services is substantially harder than debugging an equivalent issue in a monolith, because integration paths grow combinatorially with service count.

### *Organizational challenges.*

Di Francesco, Lago, and Malavolta's industrial survey found that microservices migrations surface significant business–IT alignment and knowledge-sharing challenges [5]. Participants reported difficulty reaching agreement on service boundaries and struggled to integrate domain experts into architectural decisions—a dynamic also described in Newman [8]. Without rigorous documentation, individual teams risk knowing only their service's narrow domain while losing visibility into end-to-end business workflows.

| Aspect | Monolithic | Microservices |
|---|---|---|
| **Deployment** | Single unit; one simple pipeline | Many services; orchestrated CI/CD |
| **Scaling** | Whole app, typically vertical | Per service, fine-grained |
| **Coupling** | Shared codebase and runtime | Decoupled via APIs |
| **Fault isolation** | A module fault can stop the app | Contained when fallbacks exist |
| **Complexity** | Low to start; grows in code | High in operations and testing |
| **Teams** | One or a few teams | Many autonomous teams |
| **Tech stack** | Usually uniform | Polyglot per service |
| **Cost** | Lower base overhead | Higher base; can pay off at scale |
| **Best fit** | Smaller or well-known domains | Large, high-traffic, multi-tenant |

**TABLE I. Key trade-offs between monolithic and microservices architectures.**

## CASE STUDIES AND INDUSTRY EXPERIENCE

Netflix remains the canonical microservices success story: a platform of more than a thousand microservices, frequent daily deployments, and global scale managed through internal tooling built specifically for this architecture [2], [4]. That success, however, required a deeply tech-driven organization with years of DevOps investment. As Atlassian observes, a single-product company may not need microservices at all [2].

Atlassian's Vertigo project offers a detailed migration narrative. Over a roughly two-year effort, Atlassian moved more than 100,000 customer instances of Jira and Confluence to a multi-tenant AWS architecture—completing the customer migration in about ten months with no service interruptions—while decomposing its monoliths into microservices [24]. The payoff in team autonomy and deployment velocity came only after sustained organizational commitment to orchestration tooling and cultural change [24], [2].

Amazon Prime Video's Video Quality Analysis (VQA) case offers an instructive counterpoint. The VQA team's live-stream monitoring service originally used AWS Step Functions to orchestrate detectors as individual AWS Lambda functions, with Amazon Simple Storage Service (S3) for intermediate state. High call volume drove Step Functions costs and account limits into a bottleneck, and the distributed design added overhead between detection steps. Consolidating the detectors into a single Amazon Elastic Container Service (ECS) process eliminated inter-service coordination overhead and reduced infrastructure cost by over 90 percent [11]. Su, Li, and Taibi cite this case among documented reversals from microservices to monolith [6]. It illustrates that microservices granularity should follow workload characteristics, not architectural dogma.

Faustino, Goncalves, Portela, and Rito Silva's controlled study of a stepwise monolith-to-microservices migration found that decomposing a monolith incrementally—extracting one service at a time—reduces migration risk and lets teams validate each boundary before committing to the next [18]. This establishes the modular monolith as a legitimate intermediate stage rather than a failure to commit to microservices.

## BALANCING ACT: GUIDELINES AND RECOMMENDATIONS

Given the documented trade-offs, architects should select an architecture deliberately rather than following trend.

### *Design around business domains.*

Evans' Domain-Driven Design (DDD) and bounded context pattern provide the most principled basis for service decomposition [19]. Services should map to natural business subdomains—payments, catalog, user profiles—rather than arbitrary technical layers. Di Francesco et al.'s survey confirms that practitioners who achieve clear service boundaries begin with a crystal-clear understanding of the domain [5].

### *Start simple, then extract.*

Where scale and traffic permit, a modular monolith provides a lower-complexity starting point [18]. Teams can extract services incrementally as workload characteristics demand. Faustino et al. show that incremental migration substantially reduces performance-regression risk compared with a big-bang rewrite [18].

### *When microservices fit from the outset.*

While "start with a monolith" is a sound default, certain conditions justify a microservices-first design. When a system's scale is known in advance to be very large and its traffic highly variable, fine-grained independent scaling can repay its overhead from the start, as in Netflix's streaming backend and Atlassian's multi-tenant cloud [2], [24]. Clearly bounded subdomains backed by genuine domain expertise let independent teams build in parallel without a shared codebase; Newman observes that distinct service boundaries also allow teams to match technology choices to each problem [25]. Multi-tenant SaaS products, and components with sharply different availability, performance, or security requirements, similarly benefit from isolation established early. The caveat is decisive: these benefits materialize only when service boundaries are identified correctly, and Newman cautions that getting them wrong early is expensive [25].

### *Invest in the required infrastructure.*

When microservices are warranted, invest proportionally in the supporting platform:

- **Contract testing.** As Gregor et al. demonstrate, unit tests are insufficient for catching integration faults; consumer-driven contract tests let services evolve independently while verifying interface compatibility [12], [14].
- **Service catalog and API gateway.** Atlassian's experience shows that discoverability becomes a critical operational concern once service counts exceed a few dozen [2].
- **Consistent observability standards.** Distributed tracing (OpenTelemetry), centralized log aggregation, and structured alerting should be platform-level capabilities, not per-team implementations [17].
- **Automated deployment pipelines.** Humble and Farley's Continuous Delivery describes blue-green and canary deployment patterns that let teams roll back failed service releases without affecting the wider system [20].
- **Service mesh evaluation.** For architectures exceeding roughly 20 services, Sharma and Singh recommend a service mesh to handle authentication, authorization, and observability uniformly [16].

### *Measure and be willing to adjust.*

As Su, Li, and Taibi's review demonstrates, the optimal architecture is not static [6]. Blinowski et al.'s results show that microservices outperform monoliths on throughput at high concurrency but offer little advantage—at higher cost—for smaller systems that scale vertically [7]. If cost or complexity grows disproportionate to business value, consolidation may be the right engineering decision—not a retreat from modernity.

## CONCLUSION

Both monolithic and microservices architectures serve legitimate engineering needs. We have reviewed the technical and organizational trade-offs documented in recent research and industry practice. Microservices can substantially improve scalability and team autonomy for large, high-traffic systems, but they introduce overhead in testing, operations, security, and organizational alignment that is consistently underestimated. As Felisberto concludes, architects must weigh disadvantages against specific business constraints without ideological bias [1].

We recommend the following approach: align service boundaries with business domains using DDD bounded contexts; treat the modular monolith as a valid design option for organizations that have not yet exceeded its scaling limits; adopt microservices incrementally and only where the benefits in scalability and independent deployment justify the added operational investment; and build the supporting platform—contract testing, distributed tracing, CI/CD automation, and service mesh—as a first-class concern rather than an afterthought. This balanced approach yields systems that are maintainable, cost-effective, and aligned with both current and future business needs.

### ABOUT THE AUTHOR

**Venkata Krishna Chaitanya Nuthalapati** is a Software Development Engineer and an independent researcher in distributed systems reliability and DevSecOps. He holds an M.S. in Computer Science from Georgia State University (GPA 3.96) and has presented at ICRTMD 2026. His research appears on arXiv and through IEEE.

## REFERENCES

[1] M. Felisberto, "The trade-offs between Monolithic vs. Distributed Architectures," arXiv:2405.03619 [cs.SE], May 2024.

[2] Atlassian, "Microservices vs. Monolithic Architecture," Atlassian Microservices Guide, 2023. [Online]. Available: atlassian.com/microservices/microservices-architecture/microservices-vs-monolith.

[3] Cloud Native Computing Foundation, "Cloud Native 2024: Approaching a Decade of Code, Cloud, and Change," CNCF Annual Survey, 10th ed., Apr. 2025. Available: cncf.io/reports/cncf-annual-survey-2024/.

[4] Y. Izrailevsky, S. Vlaovic, and R. Meshenberg, "Completing the Netflix Cloud Migration," Netflix Tech Blog, Jan. 2016.

[5] P. Di Francesco, P. Lago, and I. Malavolta, "Migrating Towards Microservice Architectures: An Industrial Survey," in Proc. 2018 IEEE Int'l Conf. on Software Architecture (ICSA), Seattle, WA, 2018, pp. 29–38, doi: 10.1109/ICSA.2018.00012.

[6] R. Su, X. Li, and D. Taibi, "From Microservice to Monolith: A Multivocal Literature Review," Electronics, vol. 13, no. 8, art. 1452, Apr. 2024, doi: 10.3390/electronics13081452.

[7] G. Blinowski, A. Ojdowska, and A. Przybylek, "Monolithic vs. Microservice Architecture: A Performance and Scalability Evaluation," IEEE Access, vol. 10, pp. 20357–20374, 2022, doi: 10.1109/ACCESS.2022.3152803.

[8] S. Newman, Building Microservices, 2nd ed. Sebastopol, CA: O'Reilly Media, 2021. ISBN: 978-1-492-03402-5.

[9] C. Bryar and B. Carr, Working Backwards: Insights, Stories, and Secrets from Inside Amazon. New York: St. Martin's Press, 2021. ISBN: 978-1-250-26739-9.

[10] M. Skelton and M. Pais, Team Topologies: Organizing Business and Technology Teams for Fast Flow. Portland, OR: IT Revolution, 2019. ISBN: 978-1-942788-81-2.

[11] M. Kolny, “Scaling up the Prime Video audio/video monitoring service and reducing costs by 90%,” Prime Video Tech Blog, primevideotech.com, Mar. 2023.

[12] L. Gregor, A. Hentschel, L. Kastner, and A. Pretschner, “A Taxonomy of Integration-Relevant Faults for Microservice Testing,” in Proc. 2025 IEEE ICST, 2025, pp. 138–149, doi: 10.1109/ICST62969.2025.10989000.

[13] N. Mellifera, “The Struggle for Microservice Integration Testing,” Signadot Blog, signadot.com/blog, Sep. 2024.

[14] L. Lehva, J. Makitalo, and T. Mikkonen, “Consumer-Driven Contract Tests for Microservices: A Case Study,” in SWQD 2019, LNBIP vol. 338, Springer, 2019, pp. 18–37, doi: 10.1007/978-3-030-05767-1_2.

[15] W. Li, Y. Lemieux, J. Gao, Z. Zhao, and Y. Han, “Service Mesh: Challenges, State of the Art, and Future Research Opportunities,” in Proc. 2019 IEEE SOSE, 2019, pp. 122–127, doi: 10.1109/SOSE.2019.00026.

[16] R. Sharma and Y. Singh, “Are Microservices Really the Future? Exploring the Service-Mesh Pattern with Istio,” IEEE Software, vol. 38, no. 5, pp. 91–96, 2021, doi: 10.1109/MS.2021.3080073.

[17] B. Beyer, C. Jones, J. Petoff, and N. R. Murphy, Eds., Site Reliability Engineering: How Google Runs Production Systems. Sebastopol, CA: O’Reilly, 2016. ISBN: 978-1-491-92912-4.

[18] D. Faustino, N. Goncalves, M. Portela, and A. Rito Silva, “Stepwise migration of a monolith to a microservice architecture: Performance and migration effort evaluation,” Journal of Systems and Software, vol. 210, art. 111968, 2024, doi: 10.1016/j.jss.2024.111968.

[19] E. Evans, Domain-Driven Design: Tackling Complexity in the Heart of Software. Boston, MA: Addison-Wesley, 2003. ISBN: 978-0-321-12521-7.

[20] J. Humble and D. Farley, Continuous Delivery: Reliable Software Releases through Build, Test, and Deployment Automation. Boston, MA: Addison-Wesley, 2010. ISBN: 978-0-321-60191-9.

[21] Netflix, “Conductor: a microservices orchestration engine,” GitHub repository. [Online]. Available: github.com/Netflix/conductor.

[22] Netflix, “Hystrix: a latency and fault tolerance library,” GitHub repository. [Online]. Available: github.com/Netflix/Hystrix.

[23] Netflix, “Eureka: AWS service registry for resilient mid-tier load balancing and failover,” GitHub repository. [Online]. Available: github.com/Netflix/eureka.

[24] M. Tria, “Atlassian’s Journey from Monolith to Microservices,” DeveloperWeek Global 2020, Jun. 2020. [Online]. Available: devnetwork.com.

[25] S. Newman, “Microservices for Greenfield?” samnewman.io, Apr. 2015. [Online]. Available: samnewman.io/blog/2015/04/07/microservices-for-greenfield/.